\documentclass[prd,twocolumn,preprintnumbers,amsmath,amssymb,superscriptaddress,showpacs,showkeys, nofootinbib]{revtex4-1}
\usepackage{amsmath,amsfonts,amssymb,amsthm,amstext,amscd,eucal,srcltx,mathrsfs}
\usepackage{epsfig,graphicx,bm}
\usepackage{epstopdf, epsf}
\usepackage{dcolumn}% Align table columns on decimal point
\usepackage{hyperref}
\usepackage{bm}
\usepackage{braket}
\usepackage{color}

\newcommand{\be}{\begin{eqnarray}}
\newcommand{\ee}{\end{eqnarray}}
\renewcommand{\d}{\mbox{${\rm d}$}}

\newcommand{\gn}{G_{\rm N}}
\newcommand{\R}{\mathbb{R}}
\def\eg{{\it e.g. }} 
\def\ie{{\it i.e. }}

\newcommand{\wh}[1]{\widehat{#1}}

\begin{document}
\title{MOND--like Fractional Laplacian Theory}
\author{Andrea Giusti}
\email{agiusti@ubishops.ca}
\affiliation{Department of Physics \& Astronomy, Bishop's University, 
2600 College Street, Sherbrooke Qu\'ebec, Canada J1M 1Z7}
\begin{abstract}
I provide a derivation of some characteristic effects of Milgrom's modified Newtonian dynamics (MOND) from a fractional version of Newton's theory based on the fractional Poisson equation. I employ the properties of the fractional Laplacian to investigate the features of the fundamental solution of the proposed model. The key difference between MOND and the fractional theory introduced here is that the latter is an inherently linear theory, featuring a characteristic length scale $\ell$, whilst the former is ultimately nonlinear in nature and it is characterized by an acceleration scale $a_0$. Taking advantage of the Tully-Fisher relation, as the fractional order $s$ approaches $3/2$, I then connect the length scale $\ell$, emerging from this modification of Newton's gravity, with the critical acceleration $a_0$ of MOND. Finally, implications for galaxy rotation curves of a variable-order version of the model are discussed.
\end{abstract}
\maketitle

\section{Introduction}

General Relativity (GR) and the Standard Model of particle physics have proven to be invaluable tools
for our current understanding of nature. Yet we can only account for less than 5\% of the content
of the Universe, while the rest remains mostly uncharted territory that we dub as {\em dark components}.
Over the decades, several observations \cite{Perlmutter:1997zf, Perlmutter:1998np} have proven that the 
present Universe is expanding at an accelerating pace. To explain this effect within GR 
one assumes the existence of a mysterious {\em dark energy} component pervading the Universe and accounting
for around 70\% of its energy content \cite{Amendola, Brax:2017idh}. However, since dark energy has to be included
{\em ad hoc} in Einstein's theory to reproduce the observed accelerated expansion, it is worth mentioning that 
some substantial effort has been devoted to the study of large-scale modifications of gravity aimed at reproducing this
effect while dispensing of the notion of dark energy (see \eg \cite{Capozziello:2003tk, Carroll:2003wy, 
Sotiriou:2008rp, DeFelice:2010aj, Nojiri:2010wj, Capozziello:2011et, Capozziello:2009nq, Nojiri:2017ncd}). 
Yet, all of the proposals aimed at reproducing the accelerated expansion of the Universe still do not provide an accurate explanation of many astronomical observations. For instance, if one assumes spherical symmetry then a test mass laying on a stable
Keplerian orbit around a galaxy should experience a rotational velocity $v ^2 (r) \sim \gn m(r) / r$, with $m(r)$ the total 
mass within the orbit. Observations, however, show that $v(r)$ flattens as we move away from the Galaxy Center 
(see \eg \cite{Begeman:1989kf, Zwicky:1937zza, Corbelli:1999af, Garrett:2010hd}). This effect is typically accounted for by assuming the existence of an exotic form of matter, called {\em dark matter}, such that $m(r) \sim r$ even well outside the core of the Galaxy. From the several observations of Galaxy rotation curves, as well as from measures of the mass of several Galaxy clusters, this dark component of the universe has some pretty peculiar features, namely it does not interact with electromagnetic radiation and it has an (almost) imperceptible pressure. Over the years there have been several proposals (see \eg \cite{Feng:2010gw, Duffy:2009ig, Carr:2016drx, Casadio:2018vae, Cadoni:2017evg, Cadoni:2018dnd, Tuveri:2019zor, Giusti:2019wdx}) concerning the physical nature and origin of dark matter, ranging from primordial black holes to new physics beyond the standard model, though it does not seem that we are getting any closer to a definite answer to this conundrum. An alternative to the {\em ad hoc} addition of dark matter consists in Milgrom's modified Newtonian dynamics (MOND) \cite{Milgrom:1983ca, Milgrom:1983pn, Milgrom:1983zz, Bekenstein:1984tv}, according to which Newtonian gravity is modified when the acceleration of a test mass falls below a certain threshold $a_0$, whose value is empirically determined. In detail, considering a test particle on a Keplerian orbit around a core of mass $M$, then for $a \gg a_0$ the acceleration follows the standard Newtonian theory yielding $a \simeq \gn \, M / r^2$, whilst when $a \ll a_0$ it gets modified according to $a^2 / a_0 \simeq \gn \, M / r^2$. In other words, it is possible to dispense of the notion dark matter provided that one assumes that Newton's gravity is modified at large scales, specifically leading to a transition between a short-scale $a(r) \sim 1/r^2$ behavior and $a(r) \sim 1/r$ at Galactic scales.  

	Fractional calculus \cite{FC1, FC2, FC3} is the collection of tools that allows one to extend the classical theory of calculus to the case of fractional powers of the standard integrals and derivatives. This theory then turns out to be related to the theory of weakly singular Volterra-type integro-differential operators \cite{FC4, FC5} and naturally leads to the notions of memory and nonlocality. One of the most important results of this general approach consists in a mathematically sound definition of the so-called {\em fractional Laplacian} (see \eg \cite{FL1, FL2} and references therein). 

	Here I derive the empirical asymptotic behaviors of MOND from the fundamental solution of the {\em fractional Poisson equation}
\be
\label{eq:fractionalpoisson}
(- \triangle) ^s \Phi (\bm{x}) = - 4 \, \pi \, \gn \, \ell ^{2 - 2 s} \, \rho (\bm{x}) \, ,  
\ee
where $\triangle$ denotes the standard Laplacian, $1 \leq s \lesssim 3/2$, $\Phi (\bm{x})$ the modified gravitational potential, $\rho (\bm{x})$ the matter density distribution, and $\ell$ a constant with the dimension of a length. In particular, I will consider two density profiles: the point mass $\rho (\bm{x}) = M \, \delta ^{(3)} (\bm{x})$ and the Kuzmin disk density. Then, since MOND is an inherently scale dependent effect, I discuss how this picture can be framed within the theory of variable-order fractional operators (see \eg \cite{VO}) with $s$ becoming a scale-dependent quantity $s = s(\bm{x})$.

\section{A fractional Poisson equation}
	The canonical way of approaching the problem of the mathematical definition of the {\em fractional Laplacian} requires to start from its Fourier transform. Let $f(\bm{x})$ be a function of the Schwartz class\footnote{The Schwartz class is also known as the space of rapidly decreasing functions.} on $\R ^3$, then we define the Fourier transform of $f (\bm{x})$ as 
\be
\wh{f} (\bm{k}) \equiv \mathcal{F} \left[ f (\bm{x}) \, ; \, \bm{k} \right] = 
\int _{\R ^3} e^{- i \bm{k} \cdot \bm{x}} \, f (\bm{x}) \, \d ^3 x \, ,
\ee
with $\cdot$ denoting the Euclidean scalar product. This implies that the Fourier transform of the Laplacian simply yields
\be
\mathcal{F} \left[(-\triangle) f (\bm{x}) \,  ; \, \bm{k} \right] = |\bm{k}|^2 \, \wh{f} (\bm{k}) \, ,
\ee
with $|\bm{k}|^2 \equiv \bm{k} \cdot \bm{k}$. Thus, a natural requirement for the fractional generalization of $-\triangle$ is to preserve this nice feature, \ie
\be
\label{eq:FourierFL}
\mathcal{F} \left[(-\triangle)^s f (\bm{x}) \,  ; \, \bm{k} \right] = |\bm{k}|^{2s} \, \wh{f} (\bm{k}) \, .
\ee
Note that the choice of $- \triangle$ over $\triangle$ is particularly important since the first yields a positive-definite operator, allowing one to take advantage of the {\em method of semigroups}, see \eg \cite{FL1, FL2, identity} and references therein.  Specifically, it is not hard to see that (see \eg \cite{identity}) for any $\lambda \geq 0$ one has
\be
\label{eq:lambdas}
\lambda ^s = \frac{1}{\Gamma(-s)} \int _0 ^\infty (e^{- t \, \lambda} - 1) t^{-s -1} \, \d t \, , 
\ee	
with $0 < s < 1$ and $\Gamma (z)$ denoting Euler's Gamma function, and
\be
\label{eq:lambda-s}
\lambda ^{-s} = \frac{1}{\Gamma(s)} \int _0 ^\infty e^{- t \, \lambda} \, t^{s -1} \, \d t \, ,
\ee
for any $s > 0$ and $\lambda > 0$.

Since \eqref{eq:lambdas} is not particularly helpful when trying to solve \eqref{eq:fractionalpoisson} for $s>1$, one has to rely on \eqref{eq:lambda-s}. Indeed, considering \eqref{eq:fractionalpoisson} in the Fourier domain one has
\be
\label{eq:fractionalpoissonfourier}
\wh{\Phi} (\bm{k}) = - 4 \, \pi \, \gn \, \ell ^{2 - 2 s} \, M \, |\bm{k}|^{-2s} \, ,
\ee
since, again, I am considering a configuration of the system with $\rho (\bm{x}) = M \, \delta ^{(3)} (\bm{x})$.\footnote{Rigorously speaking, it means that the solution of \eqref{eq:fractionalpoisson} corresponds to its {\em fundamental solution}.} This expression suggests the general restriction $0<s<3/2$, since otherwise $|\bm{k}|^{-2s}$ is not a tempered distribution \cite{regularity}. Eq.~\eqref{eq:fractionalpoissonfourier}, brought back to the space domain, implies that 
\be
\Phi (\bm{x}) = - 4 \, \pi \, \ell ^{2 - 2 s} \, \gn \, M \, (- \triangle) ^{-s} \delta ^{(3)} (\bm{x}) \, ,
\ee 
with $(- \triangle) ^{-s}$ denoting the {\em inverse fractional Laplacian} that, taking advantage of \eqref{eq:lambda-s} can be expressed as \cite{FL1, FL2}
\be
(- \triangle) ^{-s} f(\bm{x}) &=& \frac{1}{\Gamma(s)} \int _0 ^\infty \d t \, t^{s -1} e^{t \, \triangle} \, f(\bm{x}) \, ,  
\ee
for any $s >0$. If $0<s<3/2$, it was shown that (see \eg \cite{FL1})
\be
(- \triangle) ^{-s} \delta ^{(3)} (\bm{x}) = 
\frac{\Gamma \left(\frac{3}{2} - s \right)}{4^s \pi ^{3/2} \, \Gamma (s)} \, \frac{1}{|\bm{x}|^{3 - 2 s}} \, ,
\ee
thus leading to a potential 
\be
\Phi _s (\bm{x}) = - \frac{\Gamma \left(\frac{3}{2} - s \right)}{4^{s-1} \, \sqrt{\pi} \, \Gamma (s)} 
\left( \frac{\ell}{|\bm{x}|} \right)^{2-2s} \, \frac{\gn \, M}{|\bm{x}|} \, .
\ee
This expression clearly shows that for $s \to 1$ one easily recovers the Newtonian potential.

	Now, considering the limit for $s\to(3/2) ^-$ one is required to extend the previous argument taking into account the regularity problem that comes with this limit \cite{regularity}. Going back to \eqref{eq:fractionalpoissonfourier} and setting $s=3/2$ one finds
\be
\wh{\Phi} (\bm{k}) = - 4 \, \pi \, \gn \, \ell ^{-1} \, M \, |\bm{k}|^{-3} \, ,
\ee
which implies
\be
\label{eq:s32}
\Phi _{3/2} (\bm{x}) = - \frac{4 \pi \gn M}{\ell} 
\int_{\R^3} \frac{\d ^3 k}{(2 \pi)^3} \, \frac{e^{i \bm{k} \cdot \bm{x}}}{|\bm{k}|^3} \, 
\ee
To compute this inverse Fourier transform one can start off by denoting
\be
\nonumber \eta (\bm{k}) &=& \mathcal{F} \left[\log \left( |\bm{x}|/\ell \right) \,  ; \, \bm{k} \right] \\
&=& \int_{\R^3} \d ^3 x \, e^{- i \bm{k} \cdot \bm{x}} \, \log \left( |\bm{x}|/\ell \right) \, ,
\ee 
and by recalling that
\be
\label{Eq15}
 \triangle \log \left( |\bm{x}| / \ell \right) = \frac{1}{|\bm{x}|^2} \, .
\ee
On the one hand, taking the Fourier transform of Eq.~\eqref{Eq15} one finds
\be
\label{eq:FT-1}
\nonumber \mathcal{F} \big[ \triangle \log \left( |\bm{x}| / \ell \right) ; \bm{k} \big] &=&
\int_{\R^3} \d ^3 x \, \frac{e^{-i \bm{k} \cdot \bm{x}}}{|\bm{x}|^2} \\
\nonumber &=& \frac{4 \pi}{|\bm{k}|} \int _0 ^\infty \frac{\sin (x)}{x} \, \d x \\
&\overset{{\rm P.V.}}{=}& \frac{2 \, \pi^2}{|\bm{k}|} \, ,
\ee
where with ${\rm P.V.}$ it is understood that we are taking the principal value of the Dirichlet integral. 
On the other hand, taking advantage of the properties of the Fourier transform one finds
\be
\label{eq:FT-2}
\mathcal{F} \big[ \triangle \log \left( |\bm{x}| / \ell \right) ; \bm{k} \big] = - |\bm{k}| ^2 \, \eta (\bm{k}) \, .
\ee
Then, putting together \eqref{eq:FT-1} and \eqref{eq:FT-2} one concludes that
\be
\eta(\bm{k}) &=& \mathcal{F} \left[\log \left( |\bm{x}|/\ell \right) \,  ; \, \bm{k} \right] 
\overset{{\rm P.V.}}{=}
- \frac{2 \, \pi^2}{|\bm{k}|^3} \, .
\ee
Backtracking to \eqref{eq:s32} and exploiting the last result one finds that
\be
\Phi _{3/2} (\bm{x}) \overset{{\rm P.V.}}{=} \frac{2 \, \gn \, M}{\pi \, \ell} \, \log \left( |\bm{x}|/\ell \right) \, .
\ee

	We can therefore summarize these computations as follows:
\be
\label{eq:result}
\Phi _{s} (r)
= 
\left\{
\begin{aligned}
& - \frac{\Gamma \left(\frac{3}{2} - s \right)}{4^{s-1} \, \sqrt{\pi} \, \Gamma (s)} \left( \frac{\ell}{r} \right)^{2-2s} \frac{\gn M}{r} \, , \\
& \quad \mbox{for} \,\,  0<s <3/2 \, ,\\
& \\
& \frac{2}{\pi} \frac{\gn \, M}{\ell} \log \left( r / \ell \right) \, , \,\, \mbox{for} \,\, s = 3/2 \, ,
\end{aligned}
\right.
\ee
where the case $s=3/2$ is understood in the regularized sense discussed above.

	Note that \eqref{eq:result} simply represents the Green function for the fractional Poisson equation \eqref{eq:fractionalpoisson}, therefore the potential corresponding to a general density distribution $\rho (\bm{x})$ 
is obtained from the convolution of the latter with \eqref{eq:result}, dropping $M$. The convolution representation of the inverse fractional Laplacian is known as the {\em Riesz potential} \cite{Riesz}, which provides a map from $L^1 _{\rm loc} (\R ^3)$ onto itself for $0 < s < 3/2$.

\section{MOND--like behavior}

	The study of the fundamental solution of \eqref{eq:fractionalpoisson} suggests that this modified Newtonian theory allows one to naturally derive the transition from Newton's gravity to MOND's large-scale behavior, however MOND is never fully reproduced because of the linearity of \eqref{eq:fractionalpoisson}. These kinds of transitions are actually a rather common feature of fractional models, see \eg \cite{GianniFrancesco, Francesco-Chaos, Io-FCAA}. Additionally, this approach would suggest an inherent {\em nonlocal nature} of this MOND-like theory. Note that, differently from Milgrom's approach \cite{Milgrom:1983ca, Milgrom:1983pn, Milgrom:1983zz, Bekenstein:1984tv}, here we have introduced a constant $\ell$ with the dimension of a length, rather than $a_0$, to maintain the correct dimensions in \eqref{eq:fractionalpoisson}. But again, from $\bm{a} = - \bm{\nabla} \Phi _s$ we recover $a (r) \sim 1/r^2$ and $a (r) \sim 1/r$ for $s=1$ and $s=3/2$, respectively. Besides, from $s=3/2$ one finds that
\be
a (r) = \frac{2 \, \gn \, M}{\pi \, \ell \, r} \, ,
\ee
which, coupled to the condition $a = v^2 / r$, yields
\be
v^2 = \frac{2 \, \gn \, M}{\pi \, \ell} \, ,
\ee
which leads to the empirical {\em Tully--Fisher relation} \cite{TF}
$$
v^4 = \gn \, M \, a_0 \, ,
$$
provided that
\be
\label{eq:ell}
\ell = \frac{2}{\pi} \sqrt{\frac{\gn \, M}{a_0}} \,  ,
\ee
thus fully reconnecting the proposed fractional model with MOND's results and phenomenology.\footnote{It is also worth noting that a potential-dependency of $a_0$ was introduced in \cite{cojons} to solve problems for Galaxy clusters.}

	In order to properly equip the theory with an explicit scale-dependent behavior of $\Phi _s (r)$ one would need to replace $s$ in \eqref{eq:fractionalpoisson} with $s(\bm{x}) = s(r/\ell)$. In other words, one should convert the fractional Poisson equation of order $s$ into a variable-order fractional differential equation. This is a subject which has been largely studied (see \eg \cite{WChen} for a review) over the years paying particular attention for one-dimensional and $(1+1)$-dimensional problems. However, it is rather clear that such differential equations can only be treated numerically, whilst closed-form solutions are fairly rare and hard to find. The situation clearly worsens, both from the analytical and numerical standpoints, when one considers the case of the variable-order Laplacian. Indeed, this topic still represents a rather uncharted territory, even though a few works have started tackling this problem in the past couple of years \cite{VOL1, VOL2}.

\begin{figure}
\includegraphics[scale=0.4]{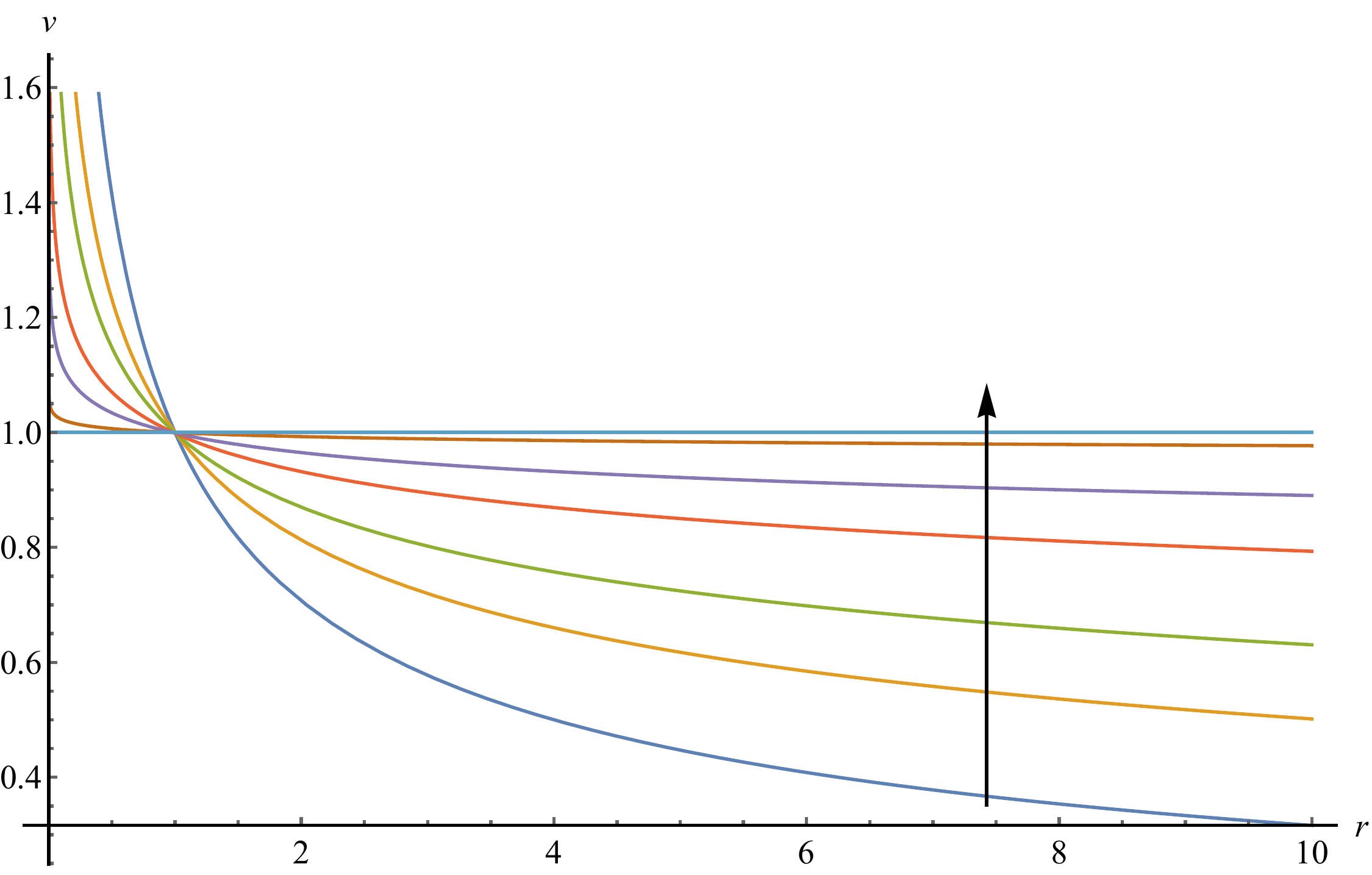}
\caption{Rotational velocity as a function of the radius $v_s(r)$ for different values of $s$. In this plot $a_0 = \gn \, M = 1$ (which implies $\ell = 2/\pi$). The arrow denotes the direction in which $s$ grows from $1$ to $3/2$.\label{fig:1}}
\end{figure}

	It is then interesting to study the behavior of the rotational velocity predicted by the proposed model as a function of $s$. Recalling that
\be
a (r) = \frac{v(r) ^2}{r} = |\bm{\nabla} \Phi_s (r)| \, ,
\ee
and that
\be
|\bm{\nabla} \Phi_s (r)|
= 
\left\{
\begin{aligned}
& \frac{4^{\frac{3}{2} - s}\Gamma \left(\frac{5}{2} - s \right)}{\sqrt{\pi} \, \Gamma (s)} \left( \frac{\ell}{r} \right)^{2-2s} \frac{\gn M}{r^2} \, , \\
& \quad \mbox{for} \,\,  0<s <3/2 \, ,\\
& \\
& \frac{2 \, \gn \, M}{\pi \, \ell \, r} \, , \,\, \mbox{for} \,\, s = 3/2 \, ,
\end{aligned}
\right.
\ee
with $\ell$ as in \eqref{eq:ell}, one finds
\be
\label{eq:vsdot}
v_s (r) = 
\left\{
\begin{aligned}
& \frac{2^{\frac{3}{2} - s}}{\sqrt[4]{\pi}} \sqrt{\frac{\Gamma \left(\frac{5}{2} - s \right)}{\Gamma (s)}}
\left( \frac{\ell}{r} \right)^{1-s} \sqrt{\frac{\gn M}{r}} \, , \\
& \quad \mbox{for} \,\,  0<s <3/2 \, ,\\
& \\
& \sqrt{\frac{2 \, \gn \, M}{\pi \, \ell}} \, , \,\, \mbox{for} \,\, s = 3/2 \, .
\end{aligned}
\right.
\ee
The velocity profiles are then plotted in Fig. \ref{fig:1}, depicting the transition from Newton's gravity to MOND. 

\section{Application to the Kuzmin disk density}
	It is now interesting to study the effect of the fractional Poisson equation \eqref{eq:fractionalpoisson} on a more realistic matter distribution. Let us consider the case of an axisymmetric system with a density function $\rho (R, z)$, in cylindrical coordinates $(R, \phi, z)$, given by
\be
\label{kuzmin-density}
\rho (R, z) = \frac{R_0 \, M}{2 \pi \, (R^2 + R_0^2)^{3/2}} \, \delta (z) \, , 
\ee
with $R_0>0$ and $[R_0] = {\rm length}$. This density profile is known in the literature as the {\em Kuzmin disk} \cite{Kuzmin, BT}.

	The simplest way to compute the corresponding potential from \eqref{eq:fractionalpoisson} requires again to employ the Fourier transform method. First, one needs to compute the momentum representation of \eqref{kuzmin-density}, {\em i.e.},
\be
\label{integralaccio}
\nonumber
\wh{\rho} (\bm{k}) &=& \int _{\R ^3} \d ^3 x \, e^{-i \bm{k} \cdot \bm{x}} \rho (\bm{x})\\
&=& \nonumber 2 \pi \int _{\R} \d z \, e^{-i k_z z} \int _0 ^\infty \rho (R, z) \, J_0 (\kappa R) \, R \, \d R \\
&=& R_0 \, M \int _0 ^\infty \frac{J_0 (\kappa R) \, R}{(R^2 + R_0^2)^{3/2}} \, \d R \, ,
\ee
with $\kappa = \sqrt{k_x^2 + k_y^2}$ and $J_0$ the standard Bessel function of the first kind and order zero. This last integral is a particular case of (see \cite{abram})
\be
\int _0 ^\infty \frac{t^{\nu + 1} \, J_\nu (\lambda \, t)}{(t^2 + z^2)^{\mu + 1}} \, \d t =
\frac{\lambda ^\mu \, z^{\nu - \mu}}{2^\mu \, \Gamma (\mu + 1)} \, K_{\nu  -\mu} (\lambda \, z) \, ,
\ee
with $\lambda > 0$, $\Re(z) > 0$, and $-1 < \Re(\nu) < 2 \, \Re (\mu) + 3/2$, and hence \eqref{integralaccio} yields
\be
\nonumber
\wh{\rho} (\bm{k}) &\equiv& \wh{\rho} (\kappa) = R_0 \, M \, \sqrt{\frac{\kappa}{2 \, R_0}} \frac{1}{\Gamma (3/2)} \, K_{-1/2} (\kappa \, R_0)\\
&=& M \, e^{- \kappa  R_0} \, ,
\ee
where $K_\nu$ is the modified Bessel function of the second kind.

	From Eq. \eqref{eq:fractionalpoisson} one has that the Fourier decomposition of the gravitational potential reads
\be
\nonumber
\wh{\Phi} (\bm{k}) &\equiv& \wh{\Phi} (\kappa, k_z)  
= - 4 \, \pi \, \gn \, \ell ^{2 - 2 s} \, \frac{\wh{\rho} (\bm{k})}{|\bm{k}|^{2s}} \\
&=& - 4 \, \pi \, \gn \, M \, \ell ^{2 - 2 s}\, \frac{e^{- \kappa R_0}}{(\kappa^2 + k_z ^2)^s} \, .
\ee
Inverting back to position space one has
\be
\nonumber
\Phi (R,z) &=& 
- 4 \, \pi \, \gn \, \ell ^{2 - 2 s}
\int_{\R^3} \frac{\d ^3 k}{(2 \pi)^3} \, e^{i \bm{k} \cdot \bm{x}} \, \frac{\wh{\rho} (\bm{k})}{|\bm{k}|^{2s}} \\
&=& \nonumber - \frac{\gn \, M \, \ell ^{2 - 2 s}}{\pi}\\
& \, & \nonumber \times \int_{\R} \d k_z \, e^{i k_z z} \int _0 ^\infty \d \kappa \, 
\frac{\kappa \,e^{-\kappa R_0} \, J_0 (\kappa R)}{(\kappa^2 + k_z ^2)^s} \\
&=& \nonumber - \frac{\gn \, M \, \ell ^{2 - 2 s}}{\pi}\\
& \, & \nonumber \times
\int _0 ^\infty \d \kappa \, \kappa \,e^{-\kappa R_0} \, J_0 (\kappa R)
\int_{\R} \d k_z \,  
\frac{e^{i k_z z}}{(\kappa^2 + k_z ^2)^s} \, .
\ee
Focusing on the case $z=0$, which is actually the most relevant one for the goals of this work, the last expression reduces to
\be
\label{kuzmin-potential-integral}
\Phi _s (R,0) &=& 
 - \frac{\gn \, M \, \ell ^{2 - 2 s}}{\sqrt{\pi}} \, \frac{\Gamma (s - 1/2)}{\Gamma (s)}\\
& \, & \nonumber \times
\int _0 ^\infty \kappa^{2-2s} \,e^{-\kappa R_0} \, J_0 (\kappa R) \, \d \kappa
\ee

If $0<s<3/2$ the integral in Eq. \eqref{kuzmin-potential-integral} converges and yields
\be
\label{kuzmin-potential-sno32}
\Phi _s (R,0) &=&
\nonumber - \frac{\gn \, M \, \ell ^{2 - 2 s}}{\sqrt{\pi} \, R_0 ^{3-2 s}} \, 
\frac{\Gamma (s - 1/2) \, \Gamma (3-2 s)}{\Gamma (s)}\\
& \, &  \times \, 
{}_2 F_1 \left(\frac{3}{2}-s, \, 2 - s, \, 1 \, ; \, - \frac{R^2}{R_0^2} \right) \, ,
\ee
with 
$
{}_2 F_1(a, \, b, \, c \, ; \,z)
$
the Gaussian hypergeometric function (see \cite{abram} for details). In particular, setting $s=1$ one finds that
\be
\Phi _1 (R,0) = - \frac{\gn \, M}{\sqrt{R^2 + R_0^2}} \, ,
\ee
which is indeed the standard Kuzmin potential evaluated on the plane of the disk \cite{Kuzmin, BT}.

If $s=3/2$ the integral in Eq. \eqref{kuzmin-potential-integral} does not converge, as in the case of the point-like source. Setting $s=3/2$ in Eq. \eqref{kuzmin-potential-integral}, one then needs to regularize the integral
\be
I(R) = \int _0 ^\infty \frac{e^{-\kappa R_0} \, J_0 (\kappa R)}{\kappa} \, \d \kappa \, .
\ee
This can be achieved by means of the {\em Hadamard regularization} (see {\em e.g.}, \cite{Riesz,Estrada})
\be
I_{\rm reg} (R) &=& \nonumber {\rm Pf} \int _0 ^\infty \frac{e^{-\kappa R_0} \, J_0 (\kappa R)}{\kappa} \, \d \kappa \\
&=& \nonumber \int _0 ^1 \frac{e^{-\kappa R_0} \, J_0 (\kappa R) - 1}{\kappa} \, \d \kappa \\
& & \quad + \int _1 ^\infty \frac{e^{-\kappa R_0} \, J_0 (\kappa R)}{\kappa} \, \d \kappa \, ,
\ee
where ${\rm Pf}$ denotes {\em Hadamard's partie finie} of the integral. In principle one could stop here, however, taking the derivative with respect to $R$ of $I_{\rm reg} (R)$ one finds
\be
\frac{\partial I_{\rm reg} (R)}{\partial R} &=& \nonumber - \int _0 ^\infty e^{- \kappa R_0} \, J_1 (\kappa R) \, \d \kappa \\
&=& \nonumber
- \frac{1}{R} \left( 1 - \frac{R_0}{\sqrt{R^2 + R_0 ^2}} \right) \\
&=& 
- \frac{\partial}{\partial R} \log \left[ 1 + \sqrt{1+ \left( \frac{R}{R_0} \right)^2} \,  \right] \, .
\ee 
This suggests that one can naively take the logarithm as the regularization of $I(R)$, since it should differ from $I_{\rm reg} (R)$ only by an integration constant. This then leads to
\be
\label{kuzmin-potential-s32}
\Phi _{3/2} (R,0) &\overset{\rm reg}{=}& \nonumber
\frac{2}{\pi} \frac{\gn \, M}{\ell} \,
\log \left[ 1 + \sqrt{1+ \left( \frac{R}{R_0} \right)^2} \,  \right] \, .\\
& &
\ee

	The the circular speed for \eqref{kuzmin-potential-sno32}, evaluated on the plane of the disk ({\em i.e.}, $z=0$), then reads
\be
\label{kuzmin-velocity-sno32}
\nonumber
v_s ^2 (R) &=& R \, \left| \bm{\nabla} \Phi _s (R,0) \right| = R \, \frac{\partial \Phi _s (R,0)}{\partial R} \\
&=& \nonumber \left( \frac{\ell}{R_0} \right)^{2-2 s} \frac{\gn M R^2}{\sqrt{\pi} \, R_0^3} \frac{(2-s) \Gamma (4-2 s) \Gamma (s-1/2)}{\Gamma (s)}\\
& \, & \times 
\, 
{}_2 F_1 \left(\frac{5}{2}-s, \, 3 - s, \, 2 \, ; \, - \frac{R^2}{R_0^2} \right) \, ,
\ee
whereas for \eqref{kuzmin-potential-s32} one finds
\be
\label{kuzmin-velocity-s32}
v_{3/2} ^2 (R) = \frac{2 \, \gn \, M}{\pi \, \ell} \left( 1 - \frac{R_0}{\sqrt{R^2 + R_0^2}} \right) \, .
\ee

\begin{figure}
\includegraphics[scale=0.4]{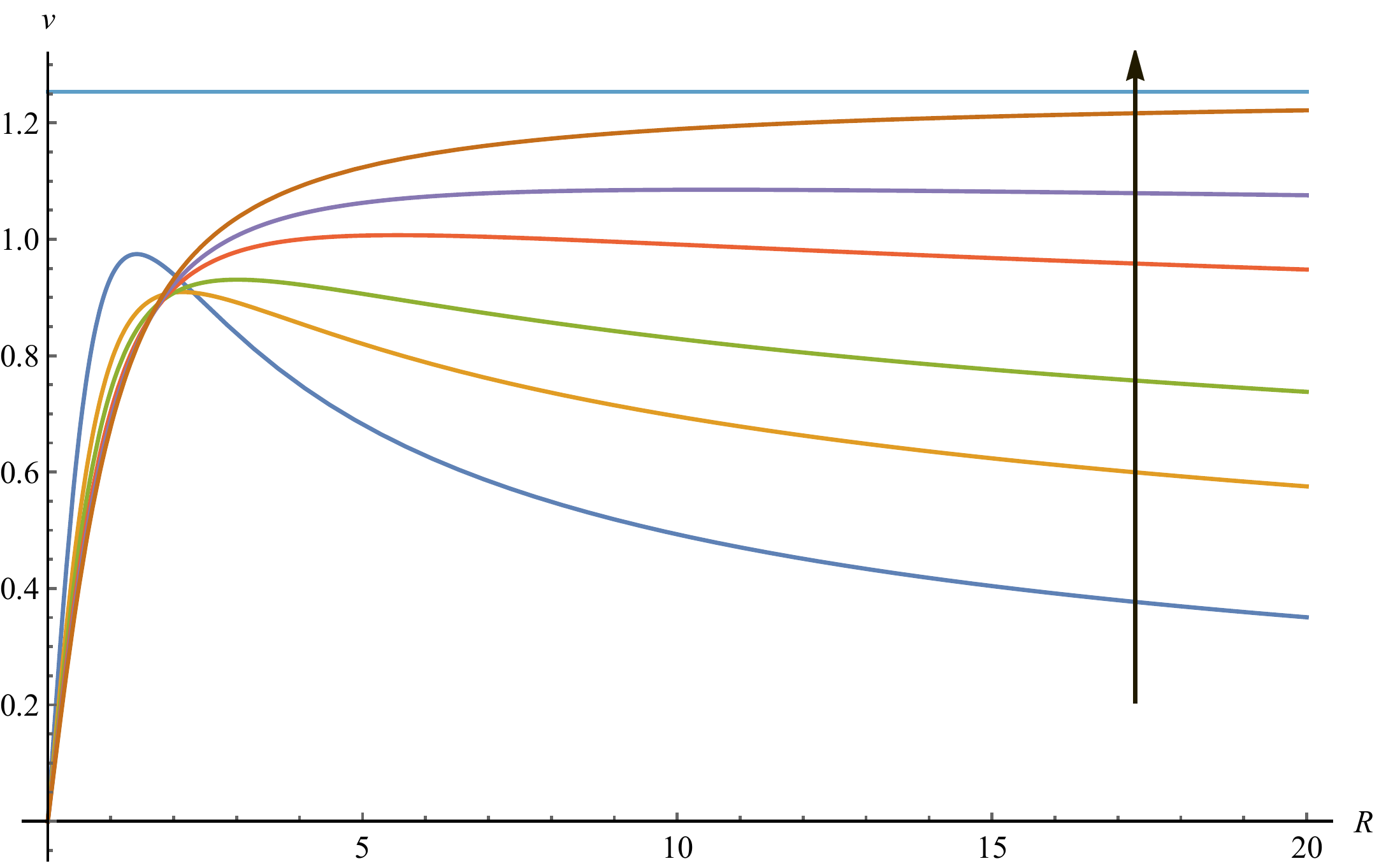}
\caption{Circular speed as a function of the radius $v_s(R)$ for different values of $s$. In this plot $a_0 = \ell = 1$ (which implies $\gn M = \pi^2 / 4$) and $R_0 = 1$. The arrow denotes the direction in which $s$ grows from $1$ to $3/2$. The horizontal line represents the asymptotic value of $v_{3/2}(R)$ as $R$ approaches infinity.\label{fig:2}}
\end{figure}

Note that
\be
v_1 ^2 (R) = \frac{\gn M \, R^2}{(R^2 + R_0^2)^{3/2}} 
\ee
is simply the circular velocity of the Kuzmin model whilst $v_{3/2} ^2 (R)$ in \eqref{kuzmin-velocity-s32} reproduces the flattening of the rotational velocity as one moves away from the Galaxy Center, with an additional $\mathcal{O}(R_0/R)$ with respect to \eqref{eq:vsdot}. Thus at large radii we recover, yet again, the identification in \eqref{eq:ell}. The velocity profiles are plotted in Fig. \ref{fig:2}.

\section{Conclusions}

	Fractional calculus has proven to be a valuable tool for studying several physical problems. Its role in fundamental physics has however been largely ignored so far, even though the past few years have finally seen the emergence of some studies pointing out its potential relevance in quantum field theory and gravity \cite{Tarasov:2018zjg, Belgacem:2017cqo, Barvinsky:2019spa, Frassino:2019yip}. In this regard, the fractional Laplacian seems to play an important role in light of its connection to the heat kernel and the Euclidean picture of nonlocal quantum field theories. Here I have derived MOND's asymptotic behaviors for the gravitational potential (and, as a consequence, for the acceleration) as the result of a fractional Newtonian theory of gravity. This model is based on the fractional Poisson equation \eqref{eq:fractionalpoisson}, where the standard Laplacian $-\triangle$ is replaced by its fractional power $(-\triangle) ^s$. Taking advantage of the method of semigroups and of the property \eqref{eq:FourierFL} I have discussed the fundamental solution of \eqref{eq:fractionalpoisson} in light of \cite{FL1,FL2}. The result is that \eqref{eq:fractionalpoisson} reduces to Newton's theory for $s=1$, whereas it reproduces MOND's large-scale behavior for $s=3/2$. This transition is, however, all but trivial since $|\bm{k}|^{-3}$ clearly does not belong to the class of tempered distributions and therefore $\Phi _{3/2} (r)$ is obtained from a regularization of the inverse Fourier transform of $|\bm{k}|^{-3}$. Then, comparing $a (r) = |\bm{\nabla} \Phi_s|$ with MOND's expression for the Tully-Fisher relation one can identify the relation between $\ell$ and $a_0$, \ie \eqref{eq:ell}. The fact that MOND predicts $a\sim 1/r$ for $a \ll a_0$ implies that $s=3/2$ for $r \gg \ell$, thus identifying a scale at which these deviations should emerge from the nonlocality of the theory. One can then infer that the proper way to fully describe the transition between the two asymptotic regimes requires to treat \eqref{eq:fractionalpoisson} as a variable-order fractional differential equation, with $s = s (r/\ell)$. Then MOND's conditions $a_0 \gg a$ and $a_0 \ll a$, which allows one to recover the flattening of the tangential velocity as one moves away from the Galaxy Center and the Newtonian force respectively, are recast as $s (r/\ell) = 3/2$ for $r \gg \ell$ and $s (r/\ell) = 1$ for $r \ll \ell$. Hence, one can frame this MOND-like linear theory as a fractional model with a variable-order spanning $1 \leq s(r/\ell) \leq 3/2$, taking proper care of the upper extreme. This nice interpretation of the model comes at a price, indeed the variable-order nature of the theory makes its analytical treatment rather subtle, if not impervious. A full numerical treatment of the variable-order counterpart of \eqref{eq:fractionalpoisson} is needed in order to constrain the functional form of $s (r/\ell)$. Note that this variability could, in principle, help explaining the fact that rotational velocities as functions of the distance from the Galaxy Center are not exactly flat and display some variability. Finally, the scale dependence of the fractional order $s (r/\ell)$ is consistent with the corpuscular interpretation of MOND \cite{Cadoni:2017evg, Cadoni:2018dnd}, which predicts that $a_0 \sim c \, H$, with $H$ denoting the Hubble parameter and $c$ being the speed of light. This is actually consistent with many observational evidences according to which $a_0 \simeq c \, H_0$ \cite{Milgrom:1983ca}. As a consequence, from \eqref{eq:ell} one finds that $\ell \sim \sqrt{\gn \,M / c \, H_0}$, providing an estimate for the critical scale for these fractional effects. 

	A natural continuation of this work would involve a precise analysis of the effects of the proposed theory for some more realistic density profiles for spherical galaxies. Particular attention should be paid to double-power-law models \cite{Zhao} since power-law effects represent a typical feature of fractional theories. The application of this MOND--like fractional Laplacian theory to other relevant astrophysical scenarios will therefore be considered elsewhere.

\section*{Acknowledgments.} 
I am grateful to R. Casadio, I. Colombaro, V. Faraoni, F. Mainardi, 
R. Garra, R. Garrappa, and Yu. Luchko for discussions.
I am supported by Bishop's University and by the Natural Sciences 
and Engineering Research Council of Canada (Grant No.~2016-03803 to V. Faraoni).
This work has also been carried out in the framework of the activities of 
the Italian National Group for Mathematical Physics (GNFM, INdAM).
\end{document}